\documentclass [11pt,a4paper]{article}
\usepackage{amsthm}

\usepackage{mathptmx}
\usepackage{amssymb}
\usepackage{amsmath}
\usepackage{eqnarray}
\usepackage{txfonts}
\usepackage[all]{xy}
\usepackage{multirow}
\usepackage{diagbox}
\def\dse#1{\vskip 0.6cm\noindent
        {\large\bf #1}
        \vskip 0.4cm}

\begin{document}

\begin{center}
\textbf{$\Large${Linear Codes over $\mathbb{F}_{q}[x]/(x^2)$ and $GR(p^2,m)$ Reaching the Griesmer Bound} }\footnote {
The work of J. Li was supported by National Natural Science Foundation of China(NSFC) under Grant 61370089,11501156 and the Anhui Provincial Natural Science Foundation under Grant 1508085SQA198. The work of A. Zhang  was supported by NSFC under Grant 11401468. The work of K. Feng was supported by the NSFC under Grant 11471178,11571007 and the Tsinghua National Lab.for Information Science and Technology. \\
J. Li is with the School of Mathematics, Hefei University of Technology, Anhui, 230001, China(e-mail: lijin\_0102@126.com )\\
A. Zhang is with the Department of Mathematical Sciences, Xi¡¯an University of Technology, Shanxi, 710048, China(e-mail:zhangaixian1008@126.com )\\
K. Feng is with the Department of Mathematical Sciences, Tsinghua University, Beijing, 100195, China(e-mail: kfeng@math.tsinghua.edu.cn)}

\end{center}

\begin{center}
\small  Jin Li, Aixian Zhang,  Keqin Feng
\end{center}


\noindent\textbf{Abstract:} We construct two series of linear codes over finite ring
$\mathbb{F}_{q}[x]/(x^2)$ and Galois ring $GR(p^2,m)$ respectively reaching the Griesmer bound.
They derive two series of codes over finite field $\mathbb{F}_{q}$ by Gray map. The first series of
codes over $\mathbb{F}_{q}$  derived from $\mathbb{F}_{q}[x]/(x^2)$ are linear and also reach the Griesmer
bound in some cases. Many of  linear codes over finite field we constructed have two Hamming (non-zero) weights.\\

\noindent\textbf{keywords:} linear code, Galois ring, homogeneous weight, Gray map, Griesmer bound.\\


\section{~~Introduction}

Let $\mathbb{F}_{q}$ be the finite field with $q$ elements, $C$ be a linear code with parameters $[n,k,d]_q$. Namely, C is a subspace of $\mathbb{F}_{q}^n$
with dimension $k=\text{dim}_{\mathbb{F}_{q}}C$ and minimum Hamming distance $d=d(C)$. The following  inequality, called the Griesmer bound, has been proved by
J.H.Griesmer[3] for $2|q$, G.Solomon and J.J.Stiffler[6] for $2\nmid q$.
 \begin{equation*}n\geq\sum_{i=0}^{k-1}\lceil\frac{d}{q^i}\rceil\tag{1.1}\end{equation*}
where $\lceil x\rceil$ denotes the smaller integer greater than or equal to $x$. Many linear codes over finite field reaching the Griesmer bound have been constructed(see [4,7] and the references there in).\\

The Griesmer bound has been generalized to linear codes over finite quasi-Frobenius ring $R$ by K.Shiromoto and L.Storme [5]. In this paper we consider $R$ being finite commutative
local ring for simplicity.\\

\noindent\textbf{Lemma 1.1([5], Theorem 2.2)} Let $R$ be a finite commutative local ring, $M$ be the (unique) maximal ideal of $R$ and $R/M=\mathbb{F}_{q}$.
Let $C$ be a linear code over $R$ with parameters $[n,k,d]$. Namely, $C$ is a free $R-$submodule of $R^n$ with rank $k$ and minimum Hamming distance $d=d(C)$.
Then
 \begin{equation*}n\geq\sum_{i=0}^{k-1}\lceil\frac{d}{q^i}\rceil\tag{1.2}\end{equation*}

There is a geometric characterization on linear codes reaching the Griesmer bound over finite field [7] and finite local ring [5].
Thus several such codes can be constructed by finite geometry (maxhypers  in projective spaces over finite field and minihypers in projective Hjelmslev spaces
over finite local ring). In this paper we construct two series of linear codes over $\mathbb{F}_{q}[x]/(x^2)$ and Galois ring $GR(p^2,m)$ reaching the
Griesmer bound (1.2) by evaluating some character sums over these finite commutative local rings(section 2 and 3). Many codes we constructed have three or
two (non-zero) Hamming weights. Moreover, in section 4, these linear codes derive two series of codes over finite field $\mathbb{F}_{q}$ by Gray map. First series of codes over
$\mathbb{F}_{q}$  derived from $\mathbb{F}_{q}[x]/(x^2)$ are linear and reach the Griesmer bound (1.1) in some cases. Many of linear codes over $\mathbb{F}_{q}$ we constructed have two Hamming weights.\\

\section{~~Linear Codes over $\mathbb{F}_{q}[x]/(x^2)$ }
\subsection{~~Basic Facts on ring  $\mathbb{F}_{q}[x]/(x^2)$}
Let $R=\mathbb{F}_{q}[x]/(x^2)=\{a+bx: a,b\in \mathbb{F}_{q}\} (x^2=0), q=p^m$. The following facts on algebraic structure of the
local commutative ring $R$ are well-known.\\

 \noindent\textbf{FACT(1)} $R=\mathbb{F}_{q}\oplus\mathbb{F}_{q}
 x$ is a two-dimensional vector space over $\mathbb{F}_{q}$. $|R|=q^2$.\\

 \noindent\textbf{FACT(2)} The unique maximal ideal of $R$ is $M=\mathbb{F}_{q}x$ and $R/M=\mathbb{F}_{q}$. The unit group (the inversible elements of $R$) is
 $$R^*=R\setminus M=\mathbb{F}_{q}^*\times(1+M) \ \ \text{(direct product)}, \ |R^*|=q(q-1).$$
We have the following isomorphism of groups
\begin{equation*}(1+M,\times)\widetilde{\longrightarrow}(\mathbb{F}_{q},+),  1+bx\longmapsto b (\in\mathbb{F}_{q})\tag{2.1}\end{equation*}
Thus the multiplicative group $1+M$ is an elementary abelian $p-$group with rank $m$, and any subgroup of $1+M=1+\mathbb{F}_{q}x$ are $1+Vx$ where
$V$ is a $\mathbb{F}_{p}$-vector subspace of $\mathbb{F}_{q}$ and  rank $(1+Vx)=\text{dim}_{\mathbb{F}_{q}}V$.\\

 \noindent\textbf{FACT(3)} Let $Q=q^s=p^{sm}$, $R^{(s)}=\mathbb{F}_{Q}[x]/(x^2)$. Then $R^{(s)}/R$ is Galois extension of rings and the Galois group is
 $Gal(R^{(s)}/R)=\langle\sigma_q\rangle$  where $\sigma_q$ is the $R-$automorphism of $R^{(s)}$ defined by
 $$\sigma_q(A+Bx)=A^q+B^qx \ \ (A,B\in\mathbb{F}_{Q})$$
 Then we have the following trace mapping
 $$Tr^{R^{(s)}}_R: R^{(s)}\longrightarrow R, \ \ A+Bx=Tr^Q_q(A)+Tr^Q_q(B)x=\sum_{i=0}^{s-1}\sigma_q^i(A+Bx) \ \ (A,B\in\mathbb{F}_{Q})$$
 which is a $R-$linear surjective mapping. For any $l\geq1$, we have
 $$Tr^{R^{(sl)}}_R(z)=Tr^{R^{(s)}}_R\circ Tr^{R^{(sl)}}_{R^{(s)}}(z) \ \ (z\in R^{(sl)})$$

\subsection{Construction of linear codes $C=C(G)$ over $R$}
Let $Q=q^s, q=p^m, s\geq2, \mathbb{F}_{Q}^*=\langle\theta\rangle$. We take a subgroup $G$ of
 $R^{(s)^*}=\mathbb{F}_{Q}^*\times(1+\mathbb{F}_{Q}x)$ which should has the following structure
 \begin{equation*}G=D\times(1+Vx)\tag{2.2}\end{equation*}
where $D=\langle\theta^e\rangle, Q-1=ef$, $V$
is a $l-$dimensional subspace of $\mathbb{F}_Q$ over $\mathbb{F}_p, 0\leq l\leq sm.$ The size of $G$ is
$n=|G|=fp^l.$
Let $G=\{g_1,\cdots,g_n\}$.

\noindent\textbf{Definition 2.1} The linear code $C(G)$ over $R$ with respect to $G$ is defined by
\begin{equation*}C=C(G)=\{c_\beta=(Tr_R^{R^{(s)}}(\beta g_1),\cdots,Tr_R^{R^{(s)}}( \beta g_n))\in R^n: \beta\in R^{(s)}\}\end{equation*}
In this section we want to determine the minimum Hamming distance $d=d(C)$. Since $C$ is linear, we have
$$d(C)=\text{min}\{w_H(c_\beta):0\neq c_\beta\in C\}$$
where $w_H(c_\beta)=\sharp\{z\in G: Tr_R^{R^{(s)}}(\beta z)\neq 0\}$ is the Hamming weight of $c_\beta$. If $w_H(c_\beta)\neq0$ for
each $0\neq\beta\in R^{(s)}$, then $c_\beta\neq c_{\beta'}$ for distinct $\beta,\beta'$ in $R^{(s)}$. In this case we have $C(G)\cong R^{(s)}$ as isomorphism of $R-$modules, $k=$rank$_RC(G)=$rank$_RR^{(s)}=s$,
$|C(G)|=|R|^s=Q^2$, and
$$d(C)=\text{min}\{w_H(c_\beta): 0\neq\beta\in R^{(s)}\}.$$
Moreover, for each $a\in R$, let $N_\beta(a)$ be the number of $a-$components in codeword $c_\beta$. Namely,
$$N_\beta(a)=\sharp\{g\in G: \ T_R^{R^{(s)}}( \beta g)=a\}$$
Then $w_H(c_\beta)=n-N_\beta(0)$ and $d(C)=n-\text{max}\{N_\beta(0):0\neq\beta\in R^{(s)}\}$.\\

In general case of $G$, the value of $N_\beta(a)$ can be expressed by Gauss sums over $R^{(s)}$ and $R$. But in this paper we consider the following particular case\\

$(*) \ \ gcd(e,\frac{Q-1}{q-1})=1$\\

Since $e|Q-1$, it is easy to see that the condition (*) is equivalent to $e|q-1$ and $gcd(e,s)=1$, where $Q=q^s$. In this case we can
determine the Hamming weight distribution
$$N_i=\sharp\{\beta\in R^{(s)}: w_H(c_\beta)=i \} \ \ (0\leq i\leq n)$$
of $C=C(G)$.\\

Under the condition (*) all computation can be done without using Gauss sums. In order to explain the reason, let us consider the case of $e=1$. Namely, let
$$\widetilde{G}=\mathbb{F}_Q^*\times(1+Vx) \ \ \ (dim_{\mathbb{F}_p}V=l)$$
and $\widetilde{C}=C(\widetilde{G})$. Each codeword in $\widetilde{C}$ is expressed as
$$\tilde{c}_\beta=(Tr_R^{R^{(s)}}(\beta z))_{z\in\widetilde{G}}\in R^{\tilde{n}}, \ \ \ \tilde{n}=|\widetilde{G}|=(Q-1)p^l=en$$
where $n=|G|$ for $G=D\times(1+Vx)$, $D=\langle\theta^e\rangle$, $\mathbb{F}_{Q}^*=\langle\theta\rangle$, and $Q-1=ef$. Since
$\mathbb{F}_{q}^*=\langle\theta^{\frac{Q-1}{q-1}}\rangle$ and $D=\langle\theta^e\rangle$, we get $\mathbb{F}_{q}^*D=\langle\theta^{gcd(e,\frac{Q-1}{q-1})}\rangle=\langle\theta\rangle=\mathbb{F}_{Q}^*$.
Namely, $\mathbb{F}_{Q}^*$ is a disjoint union of $|\mathbb{F}_{Q}^*|/|D|=e$ cosets with respect to subgroup $D$:
$$\mathbb{F}_{Q}^*=\bigcup_{i=1}^{e}(\gamma_iD) $$
and $\gamma_i(1\leq i\leq e)$ can be chosen in $\mathbb{F}_{q}^*$. Then
$$\widetilde{G}=\mathbb{F}_{Q}^*\times(1+Vx)=\bigcup_{i=1}^{e}(\gamma_iD)\times(1+Vx)=\bigcup_{i=1}^{e}\gamma_iG \ \ \text{(disjoint union)}$$
and for each $\beta\in R^{(s)^*}$,
$$c_\beta=(Tr_R^{R^{(s)}}(\beta z))_{z\in G},  \ \  \tilde{c}_{\beta}=(Tr_R^{R^{(s)}}(\beta\gamma_i z))_{z\in{G}}, \ 1\leq i\leq e$$
Since $\gamma_i\in \mathbb{F}_{q}^*$, $Tr_R^{R^{(s)}}(\beta\gamma_i z)=\gamma_iTr_R^{R^{(s)}}(\beta z)$. We get
$$w_H( \tilde{c}_\beta)=ew_H(c_\beta), \ \ N_{ei}(\widetilde{C})=N_i(C) \ \ (0\leq i\leq n)$$
and
\begin{equation*}d(\widetilde{C})=\text{min}\{w_H(\tilde{c}_\beta): \beta\in R^{(s)}\}=e\cdot\text{min}\{w_H({c}_\beta): \beta\in R^{(s)}\}=e\cdot d(C)
\tag{2.2}\end{equation*}
Therefore the computation for $C=C(G)$ is reduced to one for $\widetilde{C}=C(\widetilde{G})$, which can be done without using Gauss sums.
\subsection{ Hamming weight distribution of $C(G)$}

\noindent\textbf{Theorem 2.2} Let $q=p^m, Q=q^s (s\geq2)$, $R=\mathbb{F}_q[x]/(x^2)$, $R^{(s)}=\mathbb{F}_Q[x]/(x^2)$,
and $\mathbb{F}_Q^*=\langle\theta\rangle$. Let $C=C(G)$ be the linear code over $R$ defined by Definition 2.1,  where
$G=D\times(1+Vx)$ is a subgroup of $R^{(s)^*}$, $D=\langle\theta^e\rangle$, $Q-1=ef$ and $V$ is a $\mathbb{F}_p-$subspace of $\mathbb{F}_Q$,
dim$_{\mathbb{F}_p}V=l (0\leq l\leq sm)$. Assume that $gcd(e,\frac{Q-1}{q-1})=1$, which is equivalent to $e|q-1$ and $gcd(e,s)=1$. Then\\
\noindent(1) $C$ has parameters $[n,k,d]=[\frac{(Q-1)p^l}{e},s,\frac{Q(q-1)p^l}{eq}]$. Namely, $C$ is a free $R-$submodule of $R^n (n=\frac{(Q-1)p^l}{e})$ with
rank $k(=s)$ and minimum Hamming distance $d=\frac{Q(q-1)p^l}{eq}$. Moreover, the Hamming weight distribution of $C$ is shown in Table 1.

\begin{table*}[!hbp]\footnotesize
\begin{tabular}{|c|c|c|}
\hline
$\beta$ & $w_H(c_\beta)$& $\text{multiplicity}$\\
\hline
\multirow{2}{*}{${ \beta_1+\beta_2x\in R^{(s)^*},\atop (\beta_1\in\mathbb{F}_Q^*, \beta_2\in \mathbb{F}_Q)}$}
 &$\frac{Q(q^2-1)}{eq^2}p^l, \text{if} \ \beta_2\beta_1^{-1}\notin V+\mathbb{F}_q$&$(Q-1)(Q-|V+\mathbb{F}_q|)$\\
\cline{2-3} &$\frac{Q(q^2-1)}{eq^2}p^l-\frac{Q(q-1)}{eq^2}|V\cap\mathbb{F}_q|, \text{if} \ \beta_2\beta_1^{-1}\in V+\mathbb{F}_q$&$(Q-1)|V+\mathbb{F}_q|$\\
\hline
$\beta_2x \ (\beta_2\in\mathbb{F}_Q^*)$ & $\frac{Q(q-1)}{eq}p^l$&$Q-1$\\
\hline
0&0&1\\
\hline
\end{tabular}\caption{ The Hamming weight distribution of $C(G)$ }
\end{table*}

\noindent(2) $C$ reaches the Griesmer bound (1.2).\\

\par\noindent\textbf{Proof}. (1). By formulas (2.2) we need to consider the case $e=1$ only.
Namely, it is enough to consider $C=C(G)$ where
$$G=\mathbb{F}_Q^*\times(1+Vx), \ n=|G|=(Q-1)p^l.$$
Now we compute the values  of $w_H(c(\beta))=n-N_\beta(0)$
for each $0\neq\beta\in R^{(s)}$,\\

(\uppercase\expandafter{\romannumeral1}). Assume that $\beta=\beta_1+\beta_2x\in R^{(s)^*} \ (\beta_1\in\mathbb{F}_Q^*, \beta_2\in \mathbb{F}_Q)$.  Then for $z=z_1(1+vx)=z_1+z_1vx \ (z_1\in\mathbb{F}_Q^*, v\in V)$,
$$Tr^{R^{(s)}}_R(\beta z)=Tr^{R^{(s)}}_R(\beta_1z_1+(\beta_1z_1v+\beta_2z_1)x)=Tr^Q_q(\beta_1z_1)+Tr^Q_q((\beta_1v+\beta_2)z_1)x$$
Therefore for $a=a_1+a_2x \ (a_1,a_2\in\mathbb{F}_q)$,
\begin{align*}N_\beta(a)&=\sharp\{z_1\in\mathbb{F}_Q^*,v\in V:Tr^Q_q(\beta_1z_1)=a_1, Tr^Q_q((\beta_1v+\beta_2)z_1)=a_2\}\\
&=\sharp\{(w,v)\in\mathbb{F}_Q^*\times V:Tr^Q_q(w)=a_1, Tr^Q_q(w(v+\beta_1^{-1}\beta_2))=a_2\}\tag{2.3}
\end{align*}
where $w=\beta_1z_1$. Then we get
\begin{align*}N_\beta(0)&=\sharp\{(w,v)\in\mathbb{F}_Q^*\times V:Tr^Q_q(w)= Tr^Q_q(w(v+\beta_1^{-1}\beta_2))=0\}
\\&=\frac{1}{q^2}\sum_{z_1,z_2\in\mathbb{F}_q}\sum_{v\in V, w\in\mathbb{F}_Q^*}\zeta_p^{Tr^q_p[z_1Tr^Q_q(w)+z_2Tr^Q_q(w(v+\beta_1^{-1}\beta_2))]}
\\&=\frac{1}{q^2}(\sum_{z_1,z_2\in\mathbb{F}_q}\sum_{v\in V, w\in\mathbb{F}_Q}\zeta_p^{Tr^Q_p[w(z_2(v+\beta_1^{-1}\beta_2)+z_1)]}-q^2p^l)\\
&=-p^l+\frac{Q}{q^2}|\{z_1,z_2\in\mathbb{F}_q, v\in V: z_1+z_2(v+\beta_1^{-1}\beta_2)=0\}|\\
&=-p^l+\frac{Q}{q^2}[p^l+|\{z_1\in\mathbb{F}_q, z_2\in\mathbb{F}_q^*, v\in V: z_1+z_2(v+\beta_1^{-1}\beta_2)=0\}|]\\
&=\frac{(Q-q^2)p^l}{q^2}+\frac{Q(q-1)}{q^2}|\{z\in\mathbb{F}_q,  v\in V: z+v+\beta_1^{-1}\beta_2=0\}|
\end{align*}
The equation $z+v+\beta_1^{-1}\beta_2=0$ has solution $(z,v)\in\mathbb{F}_q\times V$ if and only if
$\beta_1^{-1}\beta_2\in\mathbb{F}_q+ V$. On the other hand, if $\beta_1^{-1}\beta_2\in\mathbb{F}_q+ V$, then
two solutions
$(z_1,v_1)$ and $(z_2,v_2)$ imply $v_1-v_2=z_2-z_1\in\mathbb{F}_q\cap V$. Thus the number of
solutions of $z+v+\beta_1^{-1}\beta_2=0 (z\in\mathbb{F}_q,  v\in V)$ is $|\mathbb{F}_q\cap V|$. Therefore
\begin{equation*}N_\beta(0)=\begin{cases}
\frac{(Q-q^2)}{q^2}p^l, &\text{if} \  \beta_2\beta_1^{-1}\notin V+\mathbb{F}_q\cr
\frac{(Q-q^2)}{q^2}p^l+\frac{Q(q-1)}{q^2}|V\cap\mathbb{F}_q|, &\text{otherwise.} \tag{2.4} \end{cases}\end{equation*}
and
\begin{align*}w_H(c_\beta)&=n-N_\beta(0)=(Q-1)p^l-\frac{(Q-q^2)p^l}{q^2}-
\begin{cases}
0, &\text{if} \  \beta_2\beta_1^{-1}\notin V+\mathbb{F}_q\cr
\frac{Q(q-1)}{q^2}|V\cap\mathbb{F}_q|, &\text{otherwise.}\end{cases}\\
&=\begin{cases}
\frac{Q(q^2-1)p^l}{q^2}, &\text{if} \  \beta_2\beta_1^{-1}\notin V+\mathbb{F}_q\cr
\frac{Q(q-1)}{q^2}[(q+1)p^l-|V\cap\mathbb{F}_q|], &\text{otherwise.}\end{cases}\end{align*}
Moreover, the multiplicity of these two weights are
$$|\{(\beta_1,\beta_2)\in\mathbb{F}_Q^*\times\mathbb{F}_Q: \beta_2\beta_1^{-1}\in V+\mathbb{F}_q\}|=(Q-1)|\{\beta_2\in\mathbb{F}_Q: \beta_2\in V+\mathbb{F}_q\}|=(Q-1)| V+\mathbb{F}_q|$$
and
$$|\{(\beta_1,\beta_2)\in\mathbb{F}_Q^*\times\mathbb{F}_Q: \beta_2\beta_1^{-1}\notin V+\mathbb{F}_q\}|=(Q-1)(Q-| V+\mathbb{F}_q|)$$
respectively.\\

(\uppercase\expandafter{\romannumeral2}). Assume that $\beta=\beta_2x \ (\beta_2\in\mathbb{F}_Q^*).$ Then
 $Tr_R^{R^{(s)}}(\beta z_1(1+vx))=Tr_q^Q(\beta_2z_1)x$. Therefore $N_\beta(a)=0$ for each $a\in R^*$, and
 $$N_\beta(0)=|\{(z_1,v)\in\mathbb{F}_Q^*\times V: Tr_q^Q(\beta_2z_1)=0\}|=(\frac{Q}{q}-1)|V|=\frac{(Q-q)p^l}{q}.$$
Therefore
$$n-N_\beta(0)=p^l(Q-1-\frac{Q-q}{q})=\frac{Q(q-1)p^l}{q}.$$
This completes the computation of Hamming weight distribution of $C$ shown in table 1.
Moreover, we have
$$\frac{Q(q-1)}{q^2}((q+1)p^l-|\mathbb{F}_q\cap V|)\geq\frac{Q(q-1)}{q^2}((q+1)p^l-p^l)=\frac{Q(q-1)}{q}p^l$$
Therefore the minimum Hamming distance of $C$ is\\

$d=\text{min}\{\frac{Q(q^2-1)p^l}{q^2},\frac{Q(q-1)p^l}{q},\frac{Q(q-1)}{q^2}((q+1)p^l-|\mathbb{F}_q\cap V|)\}=\frac{Q(q-1)}{q}p^l.$\\

Since $w_H(c(\beta))>0$ for all $0\neq\beta\in R^{(s)}$, we get $C\cong R^{(s)}$ as isomorphism of $R-$module.
The rank of $R-$module $C$ is $k=s.$ This completes the proof of $(1)$.\\

$(2)$ From $e|q-1$ and $Q=q^s$ we know that $q^{s-1}|d=\frac{Q(q-1)}{eq}p^l$. Thus
$$\sum_{i=0}^{s-1}\lceil\frac{d}{q^i}\rceil=\sum_{i=0}^{s-1}\frac{d}{q^i}=d\frac{(1-\frac{1}{Q})}{(1-\frac{1}{q})}
=\frac{Q(q-1)p^l}{eq}\cdot\frac{(Q-1)q}{(q-1)Q}=\frac{(Q-1)p^l}{e}=n$$
which means that the linear code $C(G)$ over $R$ reaches the Griesmer bound (1.2). This completes the proof of Theorem 2.2.\qed\\

\par\noindent\textbf{Remark}. $C$ has at most three (non-zero) Hamming weights
$$w_1=\frac{Q(q^2-1)}{eq^2}p^l>w_2=w_1-\frac{Q(q-1)}{eq^2}|\mathbb{F}_q\cap V|)\geq w_3=\frac{Q(q-1)}{eq}p^l$$
with multiplicity $m_1=(Q-1)(Q-|\mathbb{F}_q+ V|)$, $m_2=(Q-1)\cdot|\mathbb{F}_q+V|$ and $m_3=Q-1$ respectively.
 If $V+\mathbb{F}_q=\mathbb{F}_Q$, then $m_1=0$ and $C$ has two  Hamming weight $w_2$ and $w_3$ $(w_2>w_3)$.
 If $V\subseteq\mathbb{F}_q$, then $|V\cap\mathbb{F}_q|=p^l$, $w_2=w_3$, and $C$ has two  Hamming weight $w_1$ and $w_3$.

\section{Linear Codes over $GR(p^2,m)$}
In this section we construct a series of linear codes over Galois ring $GR(p^2,m)$ reaching the Griesmer bound.
\subsection{Algebraic structure of Galois ring $GR(p^2,m)$}
We introduce some basic facts on the Galois ring $GR(p^2,m)$, where $p$ is a prime number and $m\geq1$. For more detail we refer to Wan's book [9].\\

 \noindent\textbf{(F1)} Let $Z_{p^2}=\mathbb{{Z}}/$${p^2}\mathbb{Z}$, we have the following "module $p$" reduction homomorphism of rings
\begin{equation*}\text{(mod}p): \ Z_{p^2}\longrightarrow Z_p=\mathbb{F}_p, \ a(\text{mod}p^2)\longmapsto\overline{a}=a(\text{mod}p)\end{equation*}
This mapping can be naturally extended as a homomorphism of polynomial rings
\begin{equation*}\text{(mod}p): \ Z_{p^2}[x]\longrightarrow \mathbb{F}_p[x], f(x)=\sum_ia_ix^i\longmapsto{\overline{f}(x)}=\sum_i\overline{a}_ix^i\tag{3.1}\end{equation*}
The Galois ring $GR(p^2,m)$ is defined by  \begin{equation*}R=GR(p^2,m)=\frac{Z_{p^2}[x]}{(h(x))}\tag{3.2}\end{equation*}
where  $h(x)$ is a basic primitive polynomial of degree $m$ in $Z_{p^2}[x]$  which means that $\overline{h}(x)$ is a primitive polynomial of degree $m$
in $\mathbb{F}_p[x]$. Then the $''$module $p''$
map (3.1) induces the following homomorphism   of rings
$$\text{(mod}p): R=\frac{\ Z_{p^2}[x]}{(h(x))}\longrightarrow \overline{R}=\frac{\mathbb{F}_p[x]}{(\overline{h}(x))}=\mathbb{F}_q \ (q=p^m)$$
The kernel is the unique maximal ideal $M=pR$, $R/M=\mathbb{F}_q $ and $R^*=R\backslash M$ is the group of units. Therefore $R$ is a commutative
local ring. \\

 \noindent\textbf{(F2)} Let $\xi$ be a root of $h(x)$ in $R$. Then the order of $\xi$ is $q-1$ and $\overline{\xi}$ is a root of $\overline{h}(x)$ in $\overline{R}=\mathbb{F}_q$. By definition (3.2) of $R$, each element $\alpha$ of $R$ can be expressed uniquely as
$$\alpha=c_0+c_1\xi+\cdots+c_{r-1}\xi^{m-1} \ (c_i\in Z_{p^2})$$
and $R$ is a free $Z_{p^2}-$ module with rank $m$. On the other hand, let
$$T^*=\langle\xi\rangle=\{1,\xi,\xi^2,\cdots,\xi^{q-2}\},  T=T^*\cup\{0\}$$
Then $R=T\oplus pT$. Namely, each element $\alpha$ of $R$ can be expressed  uniquely by
$$\alpha=\alpha_1+p\alpha_2 \ (\alpha_1,\alpha_2\in T)$$
we have $M=pR=pT$. Namely, $\alpha\in M$ if and only if $\alpha_1=0$. Particularly,$\overline{T}=\mathbb{F}_q,\overline{{T}^*}=\mathbb{F}_q^*$,
$|R|=q^2=p^{2m}, |M|=q$ and $|R^*|=q(q-1)$.\\

 \noindent\textbf{(F3)} The group $R^*$ of units has the following
decomposition
 $$R^*=T^*\times(1+M) \ \ \text{(direct \ product)}$$
 where  $T^*=\langle\xi\rangle$ is a cyclic group with order $q-1$, and the multiplicative group $1+M=1+pT$ is isomorphic to the additive group $\mathbb{F}_q$
 by
\begin{gather*}(1+pT,\cdot)\widetilde{\longrightarrow}(\mathbb{F}_q,+),\ \ \
1+pc\longmapsto\overline{c} \ (c\in T) \end{gather*}

\noindent\textbf{(F4)} $GR(p^2,1)=Z_{p^2}$ and $R/{Z}_{p^2}$
is Galois extension of rings and the Galois group
is the following cyclic group
of order $m$
$$Gal(R/{Z}_{p^2})=\langle\sigma_p\rangle$$
 \noindent where $\sigma_p: R\longrightarrow R$  is a $Z_{p^2}-$automorphism of $R$ defined by
$$\sigma_p(\alpha)=\alpha_1^p+p\alpha_2^p \ \ (\text{for} \ \alpha=\alpha_1+p\alpha_2,\alpha_1,\alpha_2\in T).$$
Then we have the trace mapping:
$$Tr_{{Z}_{p^2}}^R:\ R\longrightarrow{Z}_{p^2}, \  \ Tr_{{Z}_{p^2}}^R(\alpha)=\sum_{i=0}^{m-1}\sigma^i_p(\alpha)=\big(\sum_{i=0}^{m-1}\alpha_1^{p^i}\big)+p\big(\sum_{i=0}^{m-1}\alpha_2^{p^i}\big)$$
This is a surjective homomorphism of ${Z}_{p^2}-$ modules.\\

\noindent\textbf{(F5)} Let $Q=q^s(=p^{ms}),$ and $R^{(s)}=GR(p^2,ms).$ Then $T^{(s)^*}=\langle\xi^{(s)}\rangle$ is a cyclic group with order $Q-1$, $T^{(s)}=T^{(s)^*}\cup\{0\}$ and $\overline{T^{(s)}}=\mathbb{F}_Q$,
$\overline{T^{(s)^*}}=\mathbb{F}_Q^*$, and
$$R^{(s)}=T^{(s)}\oplus pT^{(s)}=\{\alpha_1^{(s)}+p\alpha_2^{(s)}: \alpha_1^{(s)},\alpha_2^{(s)}\in T^{(s)}\}, \ |R^{(s)}|=Q^2$$
The maximal ideal of $R^{(s)}$ is $M^{(s)}=pT^{(s)}$, $|M^{(s)}|=Q$ and the group of units is
\begin{align*}R^{(s)^*}&=R^{(s)}\backslash M^{(s)}=\{\alpha_1^{(s)}+p\alpha_2^{(s)}: \alpha_1^{(s)}\in T^{(s)^*},\alpha_2^{(s)}\in T^{(s)}\}\\
&=T^{(s)^*}\times(1+M^{(s)})\ \text{(direct product)}\end{align*}
We have the following isomorphism of groups
$$(1+M^{(s)},\cdot)\widetilde{\longrightarrow}(\mathbb{F}_Q,+), \ 1+pc\longmapsto\overline{c}(\text{mod} p) \ (c\in T^{(s)})$$
The extension $R^{(s)}/R$  is Galois extension with Galois group
$$Gal(R^{(s)}/R)=\langle\sigma_q\rangle \ (\text{cyclic group of order} \ s)$$
where
$$\sigma_q(\alpha_1^{(s)}+p\alpha_2^{(s)})={ \alpha_1^{(s)}}^q+p{\alpha_2^{(s)}}^q  \  \ \ (\text{for} \  \alpha_1^{(s)},\alpha_2^{(s)}\in T^{(s)})$$
Then we have the trace mapping
$$Tr_R^{R^{(s)}}: R^{(s)}\longrightarrow R, \ \ Tr_R^{R^{(s)}}(\alpha)=\sum_{i=0}^{s-1}\sigma_q^i(\alpha)\  \ \ (\text{for} \  \alpha\in R^{(s)})$$
$Tr_R^{R^{(s)}}$ is a surjective homomorphism of $R-$ modules and the following diagram is commutative\\
\begin{equation*}\xymatrix{
 R^{(s)}\ar[d]_{ Tr_R^{R^{(s)}}} \ar[r]^{(mod p)} & \mathbb{F}_Q=\overline{R^{(s)}} \ar[d]^{{Tr}_q^Q} \\
  R \ar[d]_{Tr^R_{Z_{p^2}}} \ar[r]^{(mod p)} & \mathbb{F}_q=\overline{R} \ar[d]^{{Tr}_p^q} \\
  Z_{p^2} \ar[r]^{(mod p)} &\mathbb{F}_p =\overline{Z}_{p^2}  }\tag{3.3}\end{equation*}

\noindent\textbf{(F6)} The additive character group of $(R,+)$ is
$$\widehat{R}=\{\lambda_\beta:\beta\in R\},$$
where (for each positive integer $m,$ let $\zeta_m=e^{\frac{2\pi\sqrt{-1}}{m}}\in\mathbb{C}$)\\
 $$\lambda_\beta:\ R\longrightarrow\langle\zeta_{p^2}\rangle, \  \lambda_\beta(x)=\zeta_{p^2}^{Tr_{Z_{p^2}}^R(\beta x)} \ (x\in R)$$
By the orthogonal relation of characters, we have, for $z\in R$,
\begin{equation*}\sum_{\beta\in R}\lambda_\beta(z)=\sum_{\beta\in R}\zeta_{p^2}^{Tr_{Z_{p^2}}^R(\beta z)}
=\begin{cases}|R|, &\text{if} \  z=0\cr
0, &\text{otherwise.}
\end{cases}\tag{3.4}\end{equation*}

\subsection{Construction of linear codes $C=C(G)$ over $GR(p^2,m)$}
Let $Q=q^s, q=p^m$, $R=GR(p^2,m)$, $R^{(s)}=GR(p^2,sm)$, and
 $$R^{(s)}=T^{(s)}+pT^{(s)}, T^{(s)}=T^{(s)^*}\cup\{0\}, T^{(s)^*}=\langle\xi^{(s)}\rangle, R^{(s)^*}=T^{(s)^*}\times(1+pT^{(s)}),$$
 $$R=T+pT, T=T^*\cup\{0\}, T^*=\langle\xi\rangle, \xi=\xi^{(s)^\frac{Q-1}{q-1}}, R^*=T^*\times(1+pT).$$
Let $G$ be a subgroup of $R^{(s)^*}$ which has the following decomposition
$$G=D\times(1+pV)$$
where $D=\langle\xi^{(s)^e}\rangle$ is a subgroup of $T^{(s)^*}$, $Q-1=ef$,  $V\subseteq T^{(s)}$ and $\overline{V}$ is a
$\mathbb{F}_{p}-$subspace of $\mathbb{F}_Q=\overline{T^{(s)}}$ with  dimension $l, \ (0\leq l\leq ms).$ \\

\noindent\textbf{Definition 3.1}Let $G=\{x_1,\cdots,x_n\}$, $n=|G|=\frac{(Q-1)p^l}{e}$. The linear code $C=C(G)$ over $R=GR(p^2,m)$ is defined by
\begin{equation*}C=\{c_\beta=(Tr_R^{R^{(s)}}(\beta x_1),\cdots,Tr_R^{R^{(s)}}( \beta x_n))\in R^n: \beta\in R^{(s)}\}\end{equation*}
We also assume that $gcd(e,\frac{Q-1}{q-1})=1$ which is equivalent to $e|q-1$ and $gcd(e,s)=1$. Now we show
that the linear code $C(G)$ over $GR(p^2,m)$ reaches the Griesmer bound.\\

\noindent\textbf{Theorem 3.2} Let $G=D\times(1+pV)$ is a subgroup of $R^{(s)^*}=T^{(s)^*}\times(1+pT^{(s)})$,
where $D=\langle\xi^{(s)^e}\rangle$, $Q-1=ef$, and $\overline{V}$ is a
$\mathbb{F}_{p}-$subspace of $\mathbb{F}_Q=\overline{T^{(s)}}$ with  dimension $l$. $C=C(G)$ is the linear code over $GR(p^2,m)$ given by definition 3.1. Assume that $gcd(e,\frac{Q-1}{q-1})=1$. Then the parameters of $C$ is
$[n,k,d]=[\frac{(Q-1)p^l}{e},s,\frac{Q(q-1)p^l}{q^e}]$. Namely, $C$ is a free $R-$submodule of $R^n$ with rank $k=s$ and minimum Hamming distance $d=\frac{Q(q-1)p^l}{q^e}$. Moreover, the linear code $C$ over $R$ reaches the
Griesmer bound (1.2).

\noindent\textbf{Proof} (1) For $\beta\in R^{(s)}$, $a\in R$, we define
$N_\beta(a)$ to be the number of $a-$components of codeword $c_\beta$. By the same argument as in $R=\mathbb{F}_q[x]/(x^2)$ case, all computation can be reduced to $e=1$ case under the assumption $gcd(e,\frac{Q-1}{q-1})=1$. Thus from now on we consider $G=T^{(s)^*}\times(1+pV)$, $n=|G|=(Q-1)p^l$.\\

({\romannumeral 1}). For $\beta=\beta_1(1+p\beta_2)\in R^{(s)^*} \ (\beta_1\in T^{(s)^*}, \beta_2\in T^{(s)})$\\

Let $a\in R^*$, then
\begin{align*}N_\beta(a)
&=\frac{1}{|R|}\sum\limits_{x\in R\atop z\in G}\zeta_{p^2}^{Tr_{Z_{p^2}}^R(x(Tr_R^{R^{(s)}}(\beta z)-a))}\\&=
\frac{1}{|R|}\sum_{x\in R}\overline{\zeta}_{p^2}^{Tr_{Z_{P^2}}^R(xa)}\sum_{z\in G}\zeta_{p^2}^{Tr_{Z_{p^2}}^{R^{(s)}}(\beta zx)}
\tag{3.5}\end{align*}
\begin{equation*}N_\beta(0)=\frac{1}{|R|}\sum_{z\in G,x\in R}\zeta_{p^2}^{Tr_{Z_{P^2}}^{R^{(s)}}(\beta zx)}=\frac{1}{|R|}(\uppercase\expandafter{\romannumeral 1}+\uppercase\expandafter{\romannumeral 2})\tag{3.6}\end{equation*}
where
\begin{align*}(\uppercase\expandafter{\romannumeral 2})&=\sum\limits_{x\in pT \atop z\in G}
\zeta_{p^2}^{Tr_{Z_{p^2}}^{R^{(s)}}(\beta z x)}=\sum\limits_{x\in \mathbb{F}_q\atop z_1\in\mathbb{F}_Q^*}
{\zeta_p}^{Tr_p^Q(\overline{\beta_1} z_1 x)}|V|\\
&=p^l\sum\limits_{x\in \mathbb{F}_q\atop z_1\in\mathbb{F}_Q^*}
{\zeta_p}^{Tr_p^Q({z_1}x)}=p^l(Q-q)\end{align*}

\begin{align*}(\uppercase\expandafter{\romannumeral 1})&=\sum\limits_{x\in R^* \atop z\in G}
\zeta_{p^2}^{Tr_{Z_{p^2}}^{R^{(s)}}(\beta z x)}=\sum\limits_{x\in R^*\atop z_1\in T^{(s)^*}, \ z_2\in V}
\zeta_{p^2}^{Tr_{Z_{p^2}}^{R^{(s)}}(x\beta_1(1+p\beta_2)z_1(1+pz_2))}\\
&=|T^*|\sum\limits_{x_2\in T \atop z_1\in T^{(s)^*}, \ z_2\in V}
\zeta_{p^2}^{Tr_{Z_{p^2}}^{R^{(s)}}(z_1(1+p(\beta_2+z_2+x_2)))} \ \ (x=x_1(1+px_2), x_1\in T^*, x_2\in T)\\
&=(q-1)\sum_{z_1\in T^{(s)^*}}\zeta_{p^2}^{Tr_{Z_{p^2}}^{R^{(s)}}(z_1(1+p\beta_2))}\sum\limits_{x_2\in \mathbb{F}_q\atop  z_2\in \overline{V}}\zeta_p^{Tr^Q_p(\overline{z_1}(z_2+x_2))}\\
&=(q-1)\cdot|\overline{V}\cap\mathbb{F}_q|\sum_{z_1\in T^{(s)^*}}\zeta_{p^2}^{Tr_{Z_{p^2}}^{R^{(s)}}(z_1(1+p\beta_2))}
\sum_{x\in\overline{V}+\mathbb{F}_q}\zeta_p^{Tr^Q_p(\overline{z_1}x)}\\
&=(q-1)\cdot|\overline{V}\cap\mathbb{F}_q|\cdot|\overline{V}+\mathbb{F}_q|\cdot(\sum\limits_{z_1\in T^{(s)}\atop \overline{z_1}\in(\overline{V}+\mathbb{F}_q)^\perp}\zeta_{p^2}^{Tr_{Z_{p^2}}^{R^{(s)}}(z_1(1+p\beta_2))}-1)\\
&=(q-1)qp^l(A-1)
\end{align*}
where
\begin{equation*}A=\sum\limits_{z_1\in T^{(s)}\atop \overline{z_1}\in(\overline{V}+\mathbb{F}_q)^\perp}\zeta_{p^2}^{Tr_{Z_{p^2}}^{R^{(s)}}(z_1(1+p\beta_2))}\tag{3.7}\end{equation*}
and for any $\mathbb{F}_p-$subspace $U$ of $\mathbb{F}_Q$, $U^\perp$ is the dual subspace of $U$ defined by
 $U^\perp=\{u\in\mathbb{F}_Q:  Tr^Q_p(ux)=0 \ \text{for \ all} \ x\in U\}$.\\

 From (3.6) we get
\begin{align*}w_H(c_\beta)&=n-N_\beta(0)=(Q-1)p^l-\frac{1}{q^2}[(q-1)qp^l(A-1)+p^l(Q-q)]\\
&=\frac{(q-1)p^l}{q^2}(Q(q+1)-qA)\tag{3.8}\end{align*}
 (\romannumeral 2) For $\beta=p\beta_2$, $\beta_2\in T^{(s)^*}$.\\

 Let $z=z_1(1+pz_2)\in G \ (z_1\in T^{(s)^*}, z_2 \in V)$, then
 $$Tr_{R}^{R^{(s)}}(\beta z)=pTr_{R}^{R^{(s)}}(\beta_2 z_1)\in M$$
Therefore $N_\beta(a)=0$ for all $a\in R^*$. On the other hand,
\begin{align*}N_\beta(0)&=\frac{1}{|R|}\sum\limits_{x\in R \atop z\in G}
\zeta_{p^2}^{Tr_{Z_{p^2}}^{R}(xTr_R^{R^{(s)}}(\beta z))}
=\frac{p^l}{q}\sum\limits_{x\in T \atop z_1\in T^{(s)^*}}
\zeta_{p}^{Tr^Q_p(\overline{x\beta_2z_1})} \\
&=\frac{p^l}{q}(\sum\limits_{x\in\mathbb{F}_q  \atop z\in \mathbb{F}_Q}
\zeta_{p}^{Tr^Q_p(xz)}-q)=p^l(Qq^{-1}-1)
\end{align*}
and
\begin{equation*}w_H(c_\beta)=n-N_\beta(0)=(Q-1)p^l-p^l(Qq^{-1}-1)=\frac{Q(q-1)p^l}{q}\tag{3.9}\end{equation*}
By (3.8) we know that $A\in\mathbb{Z}$ and by (3.7) we have $A\leq|(\overline{V}+\mathbb{F}_q )^\perp|\leq|\mathbb{F}_q^\perp|=\frac{Q}{q}$. Therefore from (3.8) and (3.9) we get
\begin{align*}d=d(C)&=\text{min}\{\frac{(q-1)p^l}{q^2}(Q(q+1)-qA), \frac{Q(q-1)p^l}{q}\}\\
&=\frac{Q(q-1)p^l}{q}
\end{align*}
Moreover, from (3.8) and (3.9) we know that $w_H(c_\beta)>0$ for any $0\neq\beta\in R^{(s)}$, which
implies that $C(G)$ and $R^{(s)}$ are isomorphic as $R-$modules. Therefore $k=$rank$_RC(G)=$
rank$_R R^{(s)}=s.$ Finally, for $G=D\times(1+pV)$, $D=\langle\xi^{(s)^e}\rangle$,
$e|q-1$, $gcd(e,s)=1$. $C(G)$ is linear code over $R$ with parameters $[n,k,d]=[\frac{(Q-1)p^l}{e},s,\frac{Q(q-1)p^l}{q^e}]$. Since $R/M=\mathbb{F}_q$ and
$q^se=Qe|Q(q-1)$. We get
$$\sum_{i=0}^{s-1}\lceil\frac{d}{q^i}\rceil=\sum_{i=0}^{s-1}\frac{d}{q^i}=\frac{(Q-1)}{e}p^l=n$$
which means that the linear code $C(G)$ over $GR(p^2,m)$ reaches the Griesmer bound. This completes the proof of Theorem 3.2.\\

\noindent\textbf{Remark} The value of $A$ given by (3.7) is hard to compute in general. But for following two particular cases, $A$ and then the Hamming weight distribution of $C(G)$ can be determined.\\

\noindent\textbf{Corollary 3.3} Let $C=C(G)$ be the linear code over $R=GR(p^2,m)$ in Theorem 3.2, and
assume that $gcd(e,\frac{Q-1}{q-1})=1$.\\

(1). If $\overline{V}+\mathbb{F}_q=\mathbb{F}_Q$, then
\begin{equation*}w_H(c_\beta)
=\begin{cases}w_1=\frac{(q-1)Qp^l}{eq}, &\text{if} \  \beta=p\beta_2, \beta_2\in T^{(s)^*} \cr
w_2=w_1+\frac{(q-1)(Q-q)}{eq^2}p^l, &\text{if} \  \beta\in R^{(s)^*}
\end{cases}\end{equation*}
Particularly, $C$ has two (non-zero) Hamming weight $w_1,w_2$ with multiplicity $m_1=Q-1$ and
$m_2=Q(Q-1)$ respectively when $Q\neq q$.\\

(2). Suppose that $s=ps'$ so we have $\mathbb{F}_q\subseteq\mathbb{F}_{Q'}\subset\mathbb{F}_Q (Q'=q^{s'})$ and
$$R=GR(p^2,m)\subseteq R^{(s')}=GR(p^2,ms')\subset R^{(s)}=GR(p^2,ms)$$
If $(\overline{V}+\mathbb{F}_q)^\perp\subseteq\mathbb{F}_{Q'}$, let $W$ be the dual space of $(\overline{V}+\mathbb{F}_q)^\perp$ in $\mathbb{F}_{Q'}$. Namely,
$$W=(\overline{V}+\mathbb{F}_q)\cap\mathbb{F}_{Q'}=(\overline{V}\cap\mathbb{F}_{Q'})+\mathbb{F}_q$$
Then
\begin{equation*}w_H(c_\beta)
=\begin{cases}w_1=\frac{(q-1)Qp^l}{eq}, &\text{if} \  \beta=p\beta_2, \beta_2\in T^{(s)^*} \cr
w_2=w_1+\frac{(q-1)Qp^l}{eq^2}(1-\frac{q}{|\overline{V}+\mathbb{F}_q|}), &\text{if} \  \beta=\beta_1+p\beta_2 \ (\beta_1\in T^{(s)^*}, \ \beta_2\in T^{(s)})\atop and \ Tr^Q_{Q'}(\overline{\beta_2})\in W\cr
w_3=w_1+\frac{(q-1)Qp^l}{eq^2}=\frac{(q^2-1)Qp^l}{eq^2}, &\text{if} \  \beta=\beta_1+p\beta_2 \ (\beta_1\in T^{(s)^*}, \ \beta_2\in T^{(s)})\atop and \ Tr^Q_{Q'}(\overline{\beta_2})\notin W
\end{cases}\end{equation*}
Particularly, $C$ has three (nonzero) Hamming weights $w_1, w_2$ and $w_3$ with multiplicity $m_1=Q-1$,
$m_2=\frac{Q(Q-1)}{Q'}|W|$ and $m_3=Q(Q-1)-m_2$ respectively.\\

\noindent\textbf{Proof} From Theorem 3.2 we know that for $\beta=p\beta_2$, $\beta_2\in T^{(s)^*}$, $w_H(c_\beta)=w_1=\frac{(q-1)Qp^l}{eq}$. Now we consider $w_H(c_\beta)$ for
$$\beta=\beta_1+p\beta_2\in R^{(s)^*} \ \ (\beta_1\in T^{(s)^*}, \ \beta_2\in T^{(s)}).$$
(1). If $\overline{V}+\mathbb{F}_q=\mathbb{F}_Q$, then  $(\overline{V}+\mathbb{F}_q)^\perp=(0)$ and,
by (3.7), $A=1$. Then by (3.8) we have
$$w_H(c_\beta)=\frac{(q-1)p^l}{eq^2}(Q(q+1)-q)=w_1+\frac{(q-1)(Q-q)p^l}{eq^2}.$$
(2). If $(\overline{V}+\mathbb{F}_q)^\perp\subseteq\mathbb{F}_{Q'}, Q'=q^{s'}, s=ps'$, then by (3.7)
$A=\sum\limits_{z_1\in T^{(s')}\atop \overline{z_1}\in(\overline{V}+\mathbb{F}_q)^\perp}\zeta_{p^2}^{Tr_{Z_{p^2}}^{R^{(s)}}(z_1(1+p\beta_2))}$.
From $z_1\in T^{(s')}$ we have
$$Tr_{Z_{p^2}}^{R^{(s)}}(z_1)=Tr_{Z_{p^2}}^{R^{(s')}}(Tr^{R^{(s)}}_{R^{(s')}}(z_1))=\frac{s}{s'}Tr_{Z_{p^2}}^{R^{(s')}}(z_1)
=pTr_{Z_{p^2}}^{R^{(s')}}(z_1)$$
and
$$Tr_{Z_{p^2}}^{R^{(s)}}(z_1(1+p\beta_2))=pTr_{Z_{p^2}}^{R^{(s')}}(z_1)+pTr_{Z_{p^2}}^{R^{(s')}}(z_1Tr^{R^{(s)}}_{R^{(s')}}(\beta_2))$$
Therefore
\begin{align*}A
&=\sum_{z_1\in (\overline{V}+\mathbb{F}_q)^\perp}\zeta_{p}^{Tr_p^{Q'}(z_1(1+T^Q_{Q'}(\overline{\beta_2})))}\\
&=
\begin{cases}|(\overline{V}+\mathbb{F}_q)^\perp|, &\text{if} \  1+T^Q_{Q'}(\overline{\beta_2})\in W \cr
0, &\text{otherwise}
\end{cases}\end{align*}
From $1\in\mathbb{F}_q\subseteq W$ we know that $1+T^Q_{Q'}(\overline{\beta_2})\in W$ if and only if
$T^Q_{Q'}(\overline{\beta_2})\in W$. Then by (3.8) we get
\begin{align*}w_H(c(\beta))=
\begin{cases}w_2=\frac{(q-1)p^l}{eq^2}(Q(q+1)-q|(\overline{V}+\mathbb{F}_q)^\perp|), &\text{if} \  T^Q_{Q'}(\overline{\beta_2})\in W \cr
w_3=\frac{(q^2-1)Qp^l}{eq^2}, &\text{otherwise}
\end{cases}\end{align*}
From $|(\overline{V}+\mathbb{F}_q)^\perp|=\frac{Q}{|\overline{V}+\mathbb{F}_q|}$ we know that
$$w_2=\frac{(q-1)p^l}{eq^2}Q(q+1-\frac{q}{|(\overline{V}+\mathbb{F}_q)|})=
w_1+\frac{(q-1)Qp^l}{eq^2}(1-\frac{q}{|\overline{V}+\mathbb{F}_q|})$$
with multiplicity
$$m_2=\sharp\{\beta_1\in\mathbb{F}_Q^*, \beta_2\in\mathbb{F}_Q: T^Q_{Q'}(\overline{\beta_2})\in W \}=|\mathbb{F}_Q^*|\cdot\frac{Q}{Q'}|W|=\frac{Q(Q-1)}{Q'}|W|$$
This completes the proof of Corollary 3.3.\\

\section{Homogeneous weight and Gray map}

Linear codes over finite ring become one of hot topics in Coding theory after Hammons et al.([2], 1994) discovered that
several remarkable nonlinear binary codes are the isometric image of Gray map of linear codes over $Z_4$ with Lee weight.
The Lee weight has been generalized as homogeneous weight $w_\text{hom}$ for more general finite ring $R$([1],[8],[10] for instance), and
the isometric Gray map $\psi$ from $(R^n,w_\text{hom})$ to $(\mathbb{F}_q^{nl},w_H)$ is presented ([1] for finite chain rings and [10] for Galois ring)
with certain positive integer $l$ and finite field $\mathbb{F}_q$. In this section we will calculate the parameters of $\psi(C)$ where $C=C(G)$ is
the linear code over $R=\mathbb{F}_q[x]/(x^2)$ and $R=GR(p^2,m)$
given by Theorem 2.2 and Theorem 3.2 respectively. The computation shows that in $R=\mathbb{F}_q[x]/(x^2)$ case, the code $\psi(C)$ over $\mathbb{F}_q$ is linear with
two (nonzero) Hamming weight and reaches the Griesmer bound in some special cases. For case $R=GR(p^2,m)$, the code $\psi(C)$ over $\mathbb{F}_q$ is non-linear
and has two Hamming distance for some particular subgroup $G$ of $R^{(s)^*}$.\\

Firstly we introduce the homogeneous weight given in [1,8,10].\\

\noindent\textbf{Definition 4.1} Let $R=\mathbb{F}_q[x]/(x^2)$ or $R=GR(p^2,m)$, $q=p^m$, $M$ be the
maximal ideal of $R$ and $R^*=R\backslash M$. The homogeneous weight on $R$ is the mapping $w_\text{hom}: R\longrightarrow Z$(the ring of integers)
defined by
\begin{align*}w_\text{hom}(a)=\begin{cases}q-1, &\text{if} \ a\in R^*,  \cr q, &\text{if} \ a\in M\backslash\{0\}, \cr 0, &\text{if} \ a=0.\end{cases}
\end{align*}
More general, for any  $n\geq1$, the homogeneous weight on $R^n$ is
$w_\text{hom}: R^n\longrightarrow Z$ defined by
 $$w_\text{hom}(v)=\sum_{i=1}^{n}w_\text{hom}(v_i) \ \  \text{for} \  v=(v_1,\cdots,v_n)\in R^n.$$

If $C$ is a linear code over $R$, the minimum homogeneous distance of $C$ is
\begin{align*}d_\text{hom}(C)&=\text{min}\{w_\text{hom}(c-c'): c,c'\in C, c\neq c'\}\\
&=\text{min}\{w_\text{hom}(c): 0\neq c\in C\}\end{align*}

\noindent\textbf{Theorem 4.2} (1) Let $C=C(G)$ be the linear code over $R=\mathbb{F}_q[x]/(x^2)$ given in Theorem 2.2.
Then for $0\neq\beta=\beta_1+\beta_2x\in R^{(s)} \ (\beta_1,\beta_2\in\mathbb{F}_Q)$,
\begin{equation*}w_\text{hom}(c_\beta)
=\begin{cases}w_1=\frac{(q-1)Qp^l}{e}, &\text{if} \ \beta_1=0 \ (\beta_2\neq0), \ \text{or} \ \beta_1\neq0 \ \text{and} \  {\beta_2\beta_1^{-1}}\notin V+\mathbb{F}_q\cr
w_2=w_1-\frac{(q-1)Q}{eq}|{V}\cap\mathbb{F}_q|, &\text{if} \ \beta_1\neq0 \  \text{and} \  {\beta_2\beta_1^{-1}}\in V+\mathbb{F}_q
\end{cases}\end{equation*}
Particularly, $C$ has two (nonzero) homogeneous weights $w_1$ and $w_2$ with multiplicity $m_1=(Q-1)(Q+1-|{V}+\mathbb{F}_q|)$ and $m_2=(Q-1)|{V}+\mathbb{F}_q|$ respectively. The minimal  homogeneous distance of $C$
is $d_\text{hom}(C)=w_2$.\\

(2). Let $C=C(G)$ be the linear code over $R=GR(p^2,m)$ given in Theorem 3.2, $q=p^m$. Then for $0\neq\beta\in R^{(s)}$,
\begin{equation*}w_\text{hom}(c_\beta)
=\begin{cases}\frac{(q-1)Qp^l}{e}, &\text{if} \ \beta\in pT^{(s)^*}\cr
\frac{(q-1)p^l}{e}(Q-A), &\text{if} \  \beta=\beta_1+p\beta_2\in R^{(s)^*}
\end{cases}\end{equation*}
where $A$ is defined by (3.7) which depends on $\beta_2$.\\

\noindent\textbf{Proof}  From $\sum_{a\in M\backslash\{0\}}N_\beta(a)+\sum_{a\in R^*}N_\beta(a)+N_\beta(0)=n=\frac{(Q-1)p^l}{e}$, we get
\begin{align*}w_\text{hom}(c_\beta)&=q\sum_{a\in M\backslash\{0\}}N_\beta(a)+(q-1)\sum_{a\in R^*}N_\beta(a)\\
&=qn-\sum_{a\in R^*}N_\beta(a)-qN_\beta(0)\end{align*}
(1) Let $R=\mathbb{F}_q[x]/(x^2)$. For $\beta=\beta_2x (\beta_2\in\mathbb{F}_Q^*)$, we know that $N_\beta(a)=0$ for $a\in R^*$
and $N_\beta(0)=\frac{p^l(Q-q)}{eq}$. Therefore
$$w_\text{hom}(c_\beta)=qn-qN_\beta(0)=\frac{p^l(Q-1)q}{e}-\frac{p^l(Q-q)}{e}=\frac{p^l(q-1)Q}{e}.$$
For $\beta=\beta_1+\beta_2x\in R^{(s)^*}(\beta_1\in\mathbb{F}_Q^*, \beta_2\in\mathbb{F}_Q)$, by (2.3) we get
\begin{align*}\sum_{a\in R^*}N_\beta(a)&=\frac{1}{e}\sum_{a_1\in\mathbb{F}_q^*\atop a_2\in\mathbb{F}_q}|\{(w,v)\in\mathbb{F}_Q^*\times V: \ Tr_q^Q(w)=a_1, Tr_q^Q(w(v+\beta_1^{-1}\beta_2))=a_2\}|\\
&=\frac{1}{e}|\{(w,v)\in\mathbb{F}_Q^*\times V: \ Tr_q^Q(w)\neq0\}|=\frac{|V|}{e}\cdot\frac{Q(q-1)}{q}=\frac{Q(q-1)p^l}{eq}
\end{align*}
and by (2.4),
\begin{equation*}N_\beta(0)
=\begin{cases}\frac{(Q-q^2)p^l}{eq^2}, &\text{if} \ \beta_2\beta_1^{-1}\notin V+\mathbb{F}_q\cr
\frac{(Q-q^2)p^l}{eq^2}+\frac{Q(q-1)}{eq^2}|V\cap\mathbb{F}_q|, &\text{otherwise}
\end{cases}\end{equation*}
Therefore
\begin{align*}w_{hom}(c_\beta)&=\frac{q(Q-1)p^l}{e}-\frac{Q(q-1)p^l}{eq}-qN_\beta(0)\\&=\begin{cases}
\frac{(q-1)Qp^l}{e}, &\text{if} \ \beta_2\beta_1^{-1}\notin V+\mathbb{F}_q\cr
\frac{(q-1)Qp^l}{e}-\frac{(q-1)Q}{eq}| V\cap\mathbb{F}_q|, &\text{otherwise} \end{cases}\end{align*}
(2) Let $R=GR(p^2,m)$. For $\beta=\beta_1(1+p\beta_2)\in R^{(s)^*} (\beta_1\in T^{(s)^*}, \beta_2\in T^{(s)})$,
 by (3.5),
 \begin{align*}\sum_{a\in R^*}N_\beta(a)&=\frac{1}{|R|}\sum_{z\in G\atop x\in R}\zeta_{p^2}^{Tr_{Z_{p^2}}^{R^{(s)}}(\beta zx)}\sum_{a\in R^*}\zeta_{p^2}^{Tr_{Z_{p^2}}^{R}(xa)}\tag{4.1}
\end{align*}
If  $x=px_2\in pT^*$, then
\begin{align*}\sum_{a\in R^*}\zeta_{p^2}^{Tr_{Z_{p^2}}^{R}(xa)}&=\sum_{a\in R^*}\zeta_{p}^{Tr_p^q(\overline{ax_2})}=q\sum_{\overline{a_1}\in\mathbb{F}_q^*}\zeta_{p}^{Tr_p^q(\overline{a_1}\overline{x_2})} \ (a=a_1+pa_2)\\
&=-q
\end{align*}
If $x\in R^*$, then
$$\sum_{a\in R^*}\zeta_{p^2}^{Tr_{Z_{p^2}}^{R}(xa)}=\sum_{a\in R^*}\zeta_{p^2}^{Tr_{Z_{p^2}}^{R}(a)}=-\sum_{a\in M}\zeta_{p^2}^{Tr_{Z_{p^2}}^{R}(a)}=-\sum_{a_2\in\mathbb{F}_q}\zeta_{p}^{Tr_p^q({a_2})}=0 $$
Then by (4.1),
\begin{align*}\sum_{a\in R^*}N_\beta(a)&=\frac{1}{|R|}[|G|\cdot|R^*|-q\sum_{x_2\in T^*}\sum_{z\in G}\zeta_{p}^{Tr_p^Q(\overline{\beta zx_2})}]\\
&=\frac{(Q-1)(q-1)p^l}{eq}-\frac{1}{q}\sum_{x\in T^*}\sum_{z\in G}\zeta_p^{Tr_p^Q(\overline{x}\bar{z}\overline{\beta})}\\
&=\frac{(Q-1)(q-1)p^l}{eq}-\frac{p^l}{eq}\sum_{x\in \mathbb{F}_q^*}\sum_{z_1\in \mathbb{F}_Q^*}\zeta_p^{Tr_p^Q(xz_1)} \ (z=z_1(1+pz_2), z_2\in V)\\
&=\frac{(Q-1)(q-1)p^l}{eq}+\frac{(q-1)p^l}{eq}=\frac{Q(q-1)p^l}{eq}
\end{align*}
and by (3.6),
$$N_\beta(0)=\frac{1}{eq^2}[(q-1)qp^l(A-1)+p^l(Q-q)].$$
Therefore,
\begin{align*}w_\text{hom}(c_\beta)&=\frac{q(Q-1)p^l}{e}-\frac{Q(q-1)p^l}{eq}-\frac{p^l}{eq}[(q-1)q(A-1)+Q-q]\\
&=\frac{(q-1)p^l}{e}(Q-A)
\end{align*}
For $\beta\in pT^{(s)^*}$, we know that $N_\beta(a)=0$ for $a\in R^*$ and $N_\beta(0)=\frac{(Q-q)p^l}{eq}$.
 Thus $w_\text{hom}(c_\beta)=nq-qN_\beta(0)=\frac{Q(q-1)p^l}{e}$. This completes the proof of Theorem 4.2.\qed\\

\noindent\textbf{Remark} The linear code $C(G)$ over $\mathbb{F}_q[x]/(x^2)$ has two nonzero homogeneous weights. As
a consequence of Corollary 3.3, the linear code $C(G)$ over $GR(p^2,m)$ has two nonzero homogeneous weights in
some special cases.\\

\noindent\textbf{Corollary 4.3} Let $C=C(G)$ be the linear code over $R=GR(p^2,m)$ given in Theorem 3.2, $q=p^m$.\\

(1). If $\overline{V}+\mathbb{F}_q=\mathbb{F}_Q$, then
\begin{equation*}w_\text{hom}(c_\beta)
=\begin{cases}\frac{(q-1)Qp^l}{e}, &\text{if} \ \beta\in pT^{(s)^*}\cr
\frac{(q-1)p^l}{e}(Q-1), &\text{if} \  \beta=\beta_1+p\beta_2\in R^{(s)^*}
\end{cases}\end{equation*}

(2). Suppose that $s=ps'$, $Q'=q^{s'}$, $R^{(s')}=GR(p^2,ms')$. If $(\overline{V}+\mathbb{F}_q)^\perp\subseteq\mathbb{F}_{Q'}$, let
$W=(\overline{V}\cap\mathbb{F}_{Q'})+\mathbb{F}_q$. Then
\begin{equation*}w_\text{hom}(c_\beta)
=\begin{cases}w_1=\frac{(q-1)Qp^l}{e}, &\text{if} \ \beta\in pT^{(s)^*} \text{or} \   \beta=\beta_1+p\beta_2\in R^{(s)^*}
\text{and} \ T_{Q'}^Q(\overline{\beta_2})\notin W \cr
w_2=w_1-\frac{(q-1)p^lQ}{e|\overline{V}+\mathbb{F}_q|}, &\text{if} \  \beta=\beta_1+p\beta_2\in R^{(s)^*}
\text{and} \ T_{Q'}^Q(\overline{\beta_2})\in W
\end{cases}\end{equation*}

\noindent\textbf{Proof } (1). If $\overline{V}+\mathbb{F}_q=\mathbb{F}_Q$, then $A=1$(Corollary 3.3). The  conclusion on homogeneous weight
distribution of $C$ follows by Theorem 4.2(2).\\

(2). In this case it is shown in Corollary 3.3 that
\begin{align*}A=
\begin{cases}|(\overline{V}+\mathbb{F}_q)^\perp|=\frac{Q}{|\overline{V}+\mathbb{F}_q|}, &\text{if} \  T^Q_{Q'}(\overline{\beta_2})\in W \cr
0, &\text{otherwise}
\end{cases}\end{align*}
Then the conclusion follows by Theorem 4.2(2).\qed\\

\noindent\textbf{Definition 4.4} ([1], [10]) Let $\mathbb{F}_q=\{c_1,\cdots,c_q\}$.\\

(1). For $R=\mathbb{F}_q[x]/(x^2)$, the Gray map over $R$ is defined by
$$\psi:R\longrightarrow\mathbb{F}_q^q$$
where for $a=a_1+a_2x (a_1,a_2\in\mathbb{F}_q)$
$$\psi(a)=(c_1a_1+a_2,\cdots,c_qa_1+a_2)$$
$\psi$ is a $\mathbb{F}_q-$linear map and isometric from $(R,w_\text{hom})$ to $(\mathbb{F}_q^q,w_H)$ ([1], Theorem 1.1). Then for any $n\geq1$, we have
Gray map over $R^n$ by
$$\psi:R^n\longrightarrow\mathbb{F}_q^{qn}$$
$$\psi(v)=(\psi(v_1),\psi(v_2),\cdots,\psi(v_n)) \ \text{for} \ v=(v_1,v_2,\cdots,v_n)\in R^n$$
$\psi$ is a $\mathbb{F}_q-$linear and isometric from $(R^n,w_\text{hom})$ to $(\mathbb{F}_q^{qn},w_H)$.
Namely, for any $u,v\in R^n,$
$$w_H(\psi(u-v))=w_H(\psi(u)-\psi(v))=w_\text{hom}(u-v)$$
(2). For $R=GR(p^2,m)$, the Gray map over $R$ is defined by
$$\psi:R\longrightarrow\mathbb{F}_q^q$$
where for $a=a_1+pa_2 \ (a_1,a_2\in T, \overline{T}=\mathbb{F}_q, q=p^m)$
$$\psi(a)=(c_1\overline{a_1}+\overline{a_2},\cdots,c_q\overline{a_1}+\overline{a_2})$$
Then for any $n\geq1$, we have Gray map $\psi:R^n\longrightarrow\mathbb{F}_q^{qn}$ defined by the
same way as $R=\mathbb{F}_q[x]/(x^2)$ case. $\psi$ is isometric from $(R^n,w_\text{hom})$ to $(\mathbb{F}_q^{qn},w_H)$.
Namely, for any $u,v\in R^n$, the homogeneous distance $d_\text{hom}(u,v)=w_\text{hom}(u-v)$ between $u$ and $v$
equals to the Hamming distance $d_H(\psi(u),\psi(v))=w_H(\psi(u)-\psi(v))$ between $\psi(u)$ and $\psi(v)$.\\

\noindent\textbf{Theorem 4.5} Let $C=C(G)$ be the linear code over $R=\mathbb{F}_q[x]/(x^2)$ given in Theorem 2.2. Then $\psi(C)$
is a linear code over $\mathbb{F}_q$ with parameters $[N,k,d_H]=[\frac{(Q-1)qp^l}{e},2s,$ $\frac{Q(q-1)}{e}(p^l-\frac{|V\cap\mathbb{F}_q|}{q})]$.
Moreover, for $ 0\neq\beta=\beta_1+\beta_2x\in R^{(s)} $,
$w_H(\psi(c_\beta))=w_\text{hom}(c_\beta)$ is given by
Theorem 4.2(1). Furthermore, if ${V}+\mathbb{F}_q=\mathbb{F}_Q$,
$|V|=\frac{Q}{q}p^t$ and $\frac{(q-1)p^t}{e}<q$, then the linear code
$\psi(C)$ over $\mathbb{F}_q$ reaches the Griesmer bound.\\

\noindent\textbf{Proof} The length of $\psi(C)$ is $N=qn=\frac{(Q-1)qp^l}{e}$. The size of $\psi(C)$ is
$|R^{(s)}|=Q^2=q^{2s}$. Since $\psi$ is a $\mathbb{F}_q-$linear mapping,
the code $\psi(C)$ over $\mathbb{F}_q$ is linear and $k=\text{dim}_{\mathbb{F}_q}\psi(C)=\text{log}_qq^{2s}=2s$.
The Hamming weight distribution and the minimum Hamming distance $d_H$ of $\psi(C)$ are derived directly from Theorem 4.2(1)
and $\psi$ being isometric.\\

Finally, if ${V}+\mathbb{F}_q=\mathbb{F}_Q$, $|V|=\frac{Q}{q}p^t$, then $p^l=|V|=\frac{Q}{q}p^t$  and
$|{V}\cap\mathbb{F}_q|=\frac{p^lq}{Q}$. Thus $N=\frac{(Q-1)qp^l}{e}$ and
$$d_H=\frac{Q(q-1)p^l}{e}-\frac{Q(q-1)}{eq}|{V}\cap\mathbb{F}_q|=\frac{Q(q-1)p^l}{e}-\frac{(q-1)p^l}{e}=
\frac{Q(q-1)p^l}{e}-\frac{(q-1)Qp^t}{eq}$$
Therefore

\begin{align*}
\sum_{i=0}^{2s-1}\lceil \frac{d_H}{q^i}\rceil&=\sum_{i=0}^{s-1}\frac{(Q-1)(q-1)p^l}{eq^i}+\sum_{i=s}^{2s-1}\frac{Q(q-1)p^l}{eq^i}\\
&=\frac{(Q-1)(q-1)p^l}{e}\cdot\frac{1-\frac{1}{Q}}{1-\frac{1}{q}}+\frac{Q(q-1)p^l}{e}\cdot\frac{\frac{1}{Q}-\frac{1}{Q^2}}{1-\frac{1}{q}}\\
&=\frac{q(Q-1)^2p^l}{Qe}+\frac{q(Q-1)p^l}{Qe}=\frac{q(Q-1)p^l}{e}=N
\end{align*}
which means that $\psi(C)$ reaches the Griesmer bound. This completes the proof of Theorem 4.5. \qed\\

\noindent\textbf{Remarks} (1) The linear code $\psi(C)$  over $\mathbb{F}_q$ given by Theorem 4.5 has
two nonzero Hamming weights $w_1=\frac{Q(q-1)p^l}{e}$ and $w_2=w_1-\frac{(q-1)Q}{eq}|{V}\cap\mathbb{F}_q|$
with multiplicity $m_1=(Q-1)(Q+1-|{V}+\mathbb{F}_q|)$ and $m_2=(Q-1)|{V}+\mathbb{F}_q|$ respectively.\\

(2). Similarly, let $C=C(G)$ be the linear code over $R=GR(p^2,m)$ given in Theorem 3.2, $q=p^m$. Then
$\psi(C)$  is a (nonlinear) code over $\mathbb{F}_q$ with length $\frac{(Q-1)qp^l}{e}$ and size $|\psi(C)|=Q^2=q^{2s}$.
For $\beta,\beta'\in R^{(s)}$, the Hamming distance between $\psi(c_\beta)$ and  $\psi(c_{\beta'})$ is
$$d_H(\psi(c_\beta),\psi(c_{\beta'}))=d_\text{hom}(c_\beta-c_{\beta'})=w_\text{hom}(c_\beta-c_{\beta'}))=w_\text{hom}(c_{\beta-\beta'})$$
where $w_\text{hom}(c_\beta)$ is given in Theorem 4.2(2). Particularly by Corollary 4.3, (a). if $\overline{V}+\mathbb{F}_q=\mathbb{F}_Q$
or (b). $(\overline{V}+\mathbb{F}_q)^\perp\subseteq\mathbb{F}_{Q'} (Q'=q^{s'}, s=ps')$,  the nonlinear code $\psi(C)$  over $\mathbb{F}_q$
has two nonzero Hamming distances  $d_1$ and $d_2$ where $d_1=\frac{Q(q-1)p^l}{e}$ for both cases,  $d_2=\frac{(Q-1)(q-1)p^l}{e}$ (for case (a)),
or $d_2=\frac{Q(q-1)p^l}{e}(1-\frac{1}{|\overline{V}+\mathbb{F}_q|})$ (for case (b)).
The minimum Hamming distance of  $\psi(C)$  is $d_2$.

\dse{~~References}

\noindent [1] M.Greferath and S.E.Schmidt, Gray isometries for finite chain rings and a nonlinear ternary $(36,3^{12},5)$ code, IEEE Trans. Inform. Theory, 45(7), 1999, 2522-2524.\\

\noindent[2] A.R. Hammons Jr., P. V. Kumar, A.R. Calderbank, N.J.A. Sloane and P.
Sol$\acute{\text{e}}$, The $\mathbb{Z}_4$-linearity of Kerdock,
Preparata, Goethals, and related codes, IEEE Trans. Inform. Theory,
1994, 40: 301-319.\\

\noindent [3] J.H.Griesmer, A bound for error correcting codes, IBM J. Res. Dev., 4(5), 1960, 532-542.\\

\noindent [4] A.Klein, On codes meeting the Griesmer bound, Discrete Math.,274(1-3), 2004, 289-297.\\

\noindent[5] K.Shiromoto and L.Storme, A Griesmer bound for linear codes over finite quasi-Frobenius rings, Discrete Appli. Math., 128(1), 2003, 263-274.\\

\noindent [6] G.Solomon and J.J.Stiffler, Algebraically punctured cyclic codes, Inform. and Control, 8(2), 1965, 170-179.\\

\noindent [7] F.Tamari, A construction of some $[n,k,d]_q$ codes meeting the Griesmer bound, Discrete Math., 116(1-3), 1993, 269-287.\\

\noindent[8] J.F. Voloch and J.L.Walker, Homogeneous weights and exponential sums, Finite Fields and Their Applications, 9(3), 2003, 310-321.\\

\noindent[9]  ZheXian Wan, \emph{Lecture Notes on Finite Fields and Galois Rings}, World
Scientific, Singapore, 2003.\\

\noindent[10] B.Yildiz, A combinatorial construction of the Gray map over Galois rings,  Discrete Math., 309(10), 2009, 3408-3412.\\

\end{document}